\documentclass[preprint,showpacs,preprintnumbers,amsmath,amssymb]{revtex4}

\usepackage{graphicx}
\usepackage{dcolumn}
\usepackage{bm}
\usepackage{longtable}
\newcommand{\hpsi}[0]{\hat\psi}
\newcommand{\hpi}[0]{\hat\pi}
\newcommand{\hsigma}[0]{\hat\sigma}
\newcommand{\ha}[0]{\hat a}
\newcommand{\hc}[0]{\hat c}
\newcommand{\alal}[0]{\alpha\alpha}
\newcommand{\bebe}[0]{\beta\beta}
\newcommand{\gaga}[0]{\gamma\gamma}

\begin{document}

\preprint{APS/123-QED}

\title{Quasi-coherent state of pions in the nucleon}

\author{Masanori Morishita}
 \email{morisita@ocunp.hep.osaka-cu.ac.jp}
\author{Masaki Arima}
 \email{arima@ocunp.hep.osaka-cu.ac.jp}
\affiliation{
 Department of Physics, Osaka City University, Osaka 558-8585, Japan }

\date{\today}

\begin{abstract}

Making use of the quasi-coherent state developed by Eriksson \textit{et al.},
we can find a nucleon solution accompanied by
 the pion field with trivial topology.
We compare our approach with other related works,
and examine a coherent state description of pions in the baryon structure.
Our solution suggests a kind of nucleon resonance due to
the topological change of pion field without the usual quark excitation.
\end{abstract}


\maketitle

\section{\label{sec1} INTRODUCTION}

The pion has been accepted as one of the fundamental degrees of freedom
in the hadron physics.
The effective theories for the baryon structure,
in which the broken chiral symmetry is taken into consideration,
 provide strong coupling and nonlinearity to the pion.
Many efforts have been made to deal with these nonperturbative features
in order to reveal a pion contribution to the baryon structure.

A possible approach is
the coherent state description for the pion field.
This approach is traced back to the intermediate coupling approximation
developed and succeeded in several applications
in the particle and condensed matter physics \cite{To47,Le53}.
The idea of the coherent state has been already exploited
 in the hedgehog ansatz \cite{Sk61,Fi88}
and the coherent pair approximation (CPA) \cite{Bo81,Go88}.
The hedgehog ansatz especially is capable of
reproducing ground state properties of the nucleon \cite{Ad83,Fi88}.

However the coherent state has not been fully examined yet.
The hedgehog ansatz is not a necessary condition for the pion field
interacting with other constituents, although this ansatz is
responsible for assigning the baryon number to the Skyrmion \cite{Sk61,Wa92}.
As for the CPA, its theoretical foundation is not clear
since the spin-isospin symmetry is respected in an intuitive manner.

Furthermore, a problem concerning the isospin symmetry of the pion
 is inherent in the coherent state description. 
The two assumptions stated above are closely related with this problem.
The customary method of making the coherent state
breaks the isospin symmetry. 
Eriksson \textit{et al.} managed to obtain
the isospin-conserving coherent state for the pion field,
which is called the quasi-coherent state (QCS) \cite{Er81},
by using the Peierls-Yoccoz (PY) projection operator \cite{Pe57}.

Now we try to utilize the QCS for studying the pion properties
in a baryon without holding on the hedgehog ansatz or the CPA.
In this paper we calculate the ground state mass of the nucleon
in the linear sigma model, employing the QCS for the pion field.
Throughout our discussion we compare our approach with
 the works related with the coherent state,
i.e. the hedgehog ansatz \cite{Fi88} and the CPA \cite{Go88,Al99}.

In Sec.~\ref{sec2} we introduce the standard coherent state for the pion field
and construct the QCS in a general form.
In Sec.~\ref{sec3},
after a brief comment on the linear sigma model for the nucleon,
we make a nucleon state from the quark, sigma, and pion states.
Sec.~\ref{sec4} is devoted to
explaining the variational calculation for the nucleon mass.
There we comment on our assumption for the pion distribution function.
In Sec.~\ref{sec5}, we compare the nucleon mass calculated
by using the QCS with those obtained by the CPA and the hedgehog ansatz.
We carefully discuss similarity and difference
among these models, taking notice on the pion distribution in each model.
Finally summary and our future perspectives are given in Sec.~\ref{sec6}.
\section{\label{sec2} QUASI-COHERENT STATE}

The standard coherent state $|\Pi\rangle$ for the pion field is defined by
\begin{equation}
 \ha_i(\mathbf{k})|\Pi\rangle=f_i(\mathbf{k})|\Pi\rangle \ ,
\label{2.1}
\end{equation}
where $\ha_i(\mathbf{k})$ ($\ha_i^{\dag}(\mathbf{k})$)
 is the annihilation (creation) operator with the momentum $\mathbf{k}$
and the $i$th isospin component, satisfying the commutation relation
$\left[ \ha_i(\mathbf{k}), \ha_j^{\dag}(\mathbf{k}')\right]
 =\delta_{ij}\delta^3(\mathbf{k}-\mathbf{k}')$.
The complex function $f_i(\mathbf{k})$ 
determines the pion distribution.
Explicitly $|\Pi\rangle$ is given by (apart from the normalization factor)
\begin{equation}
  | \Pi \rangle =
 \exp\left[ \int\! d^3k \bm{f}(\mathbf{k})\cdot\hat{\bm{a}}^{\dag}(\mathbf{k})
     \right] |0\rangle \ ,
\label{2.5}
\end{equation}
where the dot represents the scalar product in the isospace.

For our later use, we introduce the spherical representation 
of $\hat{a}^{\dag}_i(\mathbf{k})$
\begin{equation}
   \hat{a}^{\dag}_{\mu lm}(k) =
    (-i)^l \int\! d\hat{k} Y_{lm}(\hat{k})\hat{a}^{\dag}_{\mu}(\mathbf{k}) \ ,
\label{2.10}
\end{equation}
where $\hat{k}$ denotes the direction of $\mathbf{k}$, 
and $\mu$ takes $\pm 1$ or 0 with the definitions
\begin{equation}
  \hat{a}_{\pm}^{\dag}(\mathbf{k})
 =\mp\frac{1}{\sqrt{2}}\left[   \hat{a}_{1}^{\dag}(\mathbf{k})
                            \pm i \hat{a}_{2}^{\dag}(\mathbf{k}) \right] \ ,
    \ \ \hat{a}_{0}^{\dag}(\mathbf{k}) = \hat{a}_{3}^{\dag}(\mathbf{k}) \ .
\label{2.15}
\end{equation}
By expanding the distribution function as 
$f_{\mu}(\mathbf{k}) = \sum (-i)^l f_{\mu lm}(k) Y_{lm}(\hat{k})$, 
Eq.~({\ref{2.5}) becomes
\begin{equation}
  |\Pi\rangle = \exp\left[
         \int\! dk \sum_{\mu lm}(-)^{\mu}f_{-\mu lm}(k)\ha^{\dag}_{\mu lm}(k)
                    \right] |0\rangle \ .
\label{2.20}
\end{equation}

Because $|\Pi\rangle$ is not an eigenstate 
both of the spin and isospin operators,
we extract an eigenstate with the isospin $(T,\mu)$
and the angular momentum $(L,M)$ from $|\Pi\rangle$
by using the Peierls-Yoccoz (PY) projection \cite{Pe57}, 
\begin{equation}
|\bm{f};T\mu\nu;LMK \rangle \equiv P^T_{\mu\nu} P^L_{MK} |\Pi\rangle \ .
\label{2.25}
\end{equation}
$P^T_{\mu\nu}$ is the PY operator for the isospin,
\begin{equation}
  P^T_{\mu\nu} = \int\! dg\ D^{T *}_{\mu\nu}(g) \hat{R}(g) \ ,
\label{2.35}
\end{equation}
where $D^{T *}_{\mu\nu}(g)$ is the rotation matrix \cite{Ro57},
and $g$ represents the Euler angle in the isospace.
The measure is defined as $\int{\rm d}g=1$.
In Eq.~(\ref{2.35}),
the factor $2T+1$ is dropped from the usual definition for brevity.
Similarly $P^L_{MK}$ is for the angular momentum,
\begin{equation}
 P^L_{MK} = \int\! dh D^{L *}_{MK}(h) R(h) \ .
\end{equation}
The Euler angle in the coordinate space is represented by $h$.
Reference \cite{Er81} first introduced Eq.~(\ref{2.35}) 
and called it the quasi-coherent state.

Note that the indices $\nu$ and $K$ are redundant in Eq.~(\ref{2.25})
because there is no `intrinsic axis' for the pion distribution.
In fact the states (\ref{2.25}) are orthogonal with respect to
the indices $(T,\mu)$ and $(L,M)$, but not to $\nu$ and $K$.
We take a linear combination \cite{Ri80}
\begin{equation}
 |\bm{f};T\mu;LM\rangle = \sum_{\nu K} C_{\nu K} |\bm{f};T\mu\nu;LMK\rangle \ ,
\label{2.40}
\end{equation}
and we consider Eq.~(\ref{2.40}) as a pion state in our calculation.
\section{\label{sec3} HAMILTONIAN AND NUCLEON STATE}

We consider the static Hamiltonian corresponding to the linear sigma model,
\begin{equation}
 \mathcal{H}=\mathcal{H}_0+\mathcal{H}_1\ .
\label{3.5}
\end{equation}
$\mathcal{H}_0$ is
\begin{equation}
 \mathcal{H}_0
 = \hpsi(\mathbf{r})^{\dag}(-i\bm{\alpha}\cdot\nabla)\hpsi(\mathbf{r})
  +\frac{1}{2}\left[ \hat{P}_{\sigma}(\mathbf{r})^2
                    +\nabla\hsigma(\mathbf{r}) ^2 \right]
  +\frac{1}{2}\left[ \hat{\bm{P}}_{\pi}(\mathbf{r})^2
                    +\nabla\hat{\bm{\pi}}(\mathbf{r})^2 \right] \ ,
\label{3.10}
\end{equation}
where $\hpsi$ is the massless quark field, and
$\hsigma$ ($\hat{P}_{\sigma}$) and $\hpi$ ($\hat{\bm{P}}_{\pi}$) are
the sigma and pion fields (their conjugate fields), respectively.
The meson-quark interaction and meson self-interaction are
included in 
\begin{eqnarray}
 {\mathcal H}_1 &=&
   G\bar{\hpsi}(\mathbf{r})
       \left[ \hsigma(\mathbf{r})
             + i\gamma_5\bm{\tau}\cdot\hat{\bm{\pi}}(\mathbf{r})
       \right] \hpsi(\mathbf{r})
\nonumber\\
 &&+\frac{\lambda^2}{4}\left[
    \hsigma(\mathbf{r})^2 + \hat{\bm{\pi}}(\mathbf{r})^2 - \nu^2
    \right]^2
    - m_{\pi}^2 f_{\pi} \left[ \hsigma(\mathbf{r}) - f_{\pi} \right]
    - \frac{m_{\pi}^4}{4\lambda^2} \ ,
\label{3.15}
\end{eqnarray}
where $G$ is the coupling constant,
 $m_{\pi}$ the pion mass, and $f_{\pi}$ the pion decay constant.
The parameter $\nu$ and the self-interaction strength $\lambda$ is related
to $m_{\pi}$, $f_{\pi}$, and the sigma mass $m_{\sigma}$ as
$\nu^2 = f_{\pi}^2 - m_{\pi}^2/\lambda^2$, 
$\lambda^2 = (m_{\sigma}^2 - m_{\pi}^2)/(2f_{\pi}^2)$.

We consider the ground state baryon in this work,
and assume that all quarks are in the lowest $s$-wave.
We expand the quark field as
\begin{equation}
 \hpsi(\mathbf{r}) = \sum_{\mu m}\left(
    \langle \mathbf{r} |\mu m \rangle \hat{d}_{\mu m} 
  + \langle \mathbf{r} |\mu m \rangle^* \hat{d}^{\dag}_{\mu m} \right)  \ ,
\label{3.25}
\end{equation}
where $\hat{d}_{\mu m}$ annihilates the quark with the isospin $\mu$ and
the spin $m$. 
The quark spinor is written as
\begin{equation}
  \langle \mathbf{r} |\mu m \rangle
 = \left( \begin{array}{c} u(r) \\
                          iv(r)\bm{\sigma}\cdot\hat{\bm{r}}
          \end{array} \right)  \chi_m \zeta_{\mu} \ ,
\label{3.30}
\end{equation}
where $\chi_m$ is the two-spinor,
and $\zeta_{\mu}$ the isospinor.

Since the quark excitation is not considered,
the sigma meson always remains in the $s$-wave
through the scalar interaction in $\mathcal{H}_1$.
And only the $p$-wave pion can interact with the quark
through the pseudoscalar interaction.
The meson fields are expanded in the spherical representation as
\begin{eqnarray}
    \hsigma(\mathbf{r})
 &=&\int_0^{\infty}\!\frac{{\rm d}k k^2}{\sqrt{\pi \omega_{\sigma}}}
     j_0(kr) \left[  \hat{c}(k) + \hat{c}^{\dag}(k)
                      \right] Y_{00}^*(\hat{r}) \ ,
\label{3.35}\\
    \hpi_{\mu}(\mathbf{r})
 &=&\int_0^{\infty}\!\frac{{\rm d}k k^2}{\sqrt{\pi \omega_{\pi}}}
    \sum_{m} j_1(kr) \left[ (-)^{\mu+m} \hat{a}_{-\mu, -m}(k) 
                            + \hat{a}^{\dag}_{\mu m}(k)
                     \right] Y_{1m}^*(\hat{r}) \ ,
\label{3.40}
\end{eqnarray}
where $\omega_{\sigma} = \sqrt{ k^2 + m_{\sigma}^2 }$,
 $\omega_{\pi} = \sqrt{ k^2 + m_{\pi}^2 }$,
and $j_l(kr)$ is the spherical Bessel function.
The annihilation operator for the $s$-wave sigma field is denoted by $\hc(k)$,
and $\ha_{\mu 1m}(k)$ is simply written as $\ha_{\mu m}(k)$.

We consider that a nucleon state is composed of
quark, pion, and sigma states as in Refs.~\cite{Fi88,Go88}.
Following the usual prescription in the constituent quark model \cite{Is79},
we make a three-quark state with the spin-isospin 1/2
denoted by $|\text{3q};N\rangle$
and that with 3/2 by $|\text{3q};\Delta\rangle$.

The standard coherent state description is used for a sigma meson state,
\begin{equation}
 |\Sigma\rangle
 = \exp\left\{ \int_0^{\infty}\! dk k^2 \left[
        \eta(k)^* \hc(k) -\eta(k) \hc^{\dag}(k) \right] \right\} |0\rangle \ ,
\label{3.45}
\end{equation}
which satisfies the standard definition,
$\hc(k)|\Sigma\rangle = \eta(k)|\Sigma\rangle$.
The complex function $\eta(k)$ represents 
the momentum distribution.

Employing the QCS for the pion state explained in the last section,
we write down the nucleon state as a linear combination
with the coefficients $\alpha$, $\beta$, and $\gamma$,
\begin{eqnarray}
  |N\rangle
 &=& \left( \alpha |{\rm 3q};N\rangle|\bm{f};00\rangle
          + \beta \left[  |{\rm 3q};N\rangle
                  \otimes |\bm{f};11\rangle \right]^{\frac{1}{2} \frac{1}{2}}
          + \gamma\left[  |{\rm 3q};\Delta\rangle
                  \otimes |\bm{f};11\rangle \right]^{\frac{1}{2} \frac{1}{2}}
     \right)|\Sigma\rangle \ ,
\label{3.60}
\end{eqnarray}
where $|\bm{f};00\rangle$ 
$(|\bm{f};11\rangle)$
means the QCS 
with $T=L=0$ ($T=L=1$).
We obtain the delta state $|\Delta\rangle$ with the spin-isospin 3/2
 by interchanging $|{\rm 3q};N\rangle$ and $|{\rm 3q};\Delta\rangle$
 in Eq.~(\ref{3.60}).
\section{\label{sec4} CALCULATION}

We calculate the expectation value of
$H=\int\! d^3 r {\mathcal H}$ by the variational method,
\begin{eqnarray}
 \delta \left\{ \langle N | H | N\rangle
   -E \langle N | N\rangle
   -12\pi\epsilon\int_0^{\infty}\! dr r^2 \left[ u(r)^2 + v(r)^2 \right]
      \right\} = 0 \ ,
\label{4.1}
\end{eqnarray}
where the Lagrange multipliers $E$ and $\epsilon$ are introduced
to normalize $|N\rangle$ and the quark wave function,
and $E$ corresponds to the nucleon mass.
We notice that the QCS (\ref{2.40}) is not normalized and its
norm depends on $f_{\mu}(\mathbf{k})$.
The variation is taken with respect to the coefficients
$\alpha$, $\beta$, and $\gamma$, and also to the quark and meson fields.
Then we obtain the energy eigenvalue equation for the coefficients
 and the differential equations for the fields. 

Here we consider the pion distribution $f_{\mu}(\mathbf{k})$.
If there is no apparent correlation between the isospin and coordinate 
spaces,
we can write the $p$-wave component of
$f_{\mu}(\mathbf{k})$ as $-i f_{\mu 0}(k) Y_{10}(\hat{k})$,
taking into account the axial symmetry of ${\cal H}$.
Further we assume that 
all the three components have the same momentum dependence for simplicity: 
$f_{\mu 0}(k) = f_{\mu}\xi(k)$,
where $f_{\mu}$ is a constant vector in the isospace
and $\xi(k)$ is a real function of $k$.
This simplest form we choose here
should be compared with other choices such as the hedgehog form.
We discuss this point further in the next section. 

We do not take the variation
with respect to $C_{\nu K}$ introduced in the QCS (9)
because these coefficients are related with $f_{\mu}$.
Let us consider, for example, the matrix element of the pion kinetic energy
$H_{\pi}=\int\!{\rm d}^3k \omega_{\pi}
 \hat{\bm{a}}^{\dag}(\mathbf{k})\cdot\hat{\bm{a}}(\mathbf{k})$
between the QCS (9),
\begin{eqnarray}
 &&\langle \bm{f};T\mu;LM| H_{\pi} |\bm{f};T\mu;LM \rangle
\nonumber\\
 &&=\sum_{\nu'\nu\lambda'\lambda} C^*_{\nu' 0}C_{\nu 0}
                  f_{\lambda'}^* f_{\lambda}
                \int\! dg dh
   D_{\nu'\nu}^{1 *}(g) D_{\lambda'\lambda}^{1 *}(g)
   D_{00}^{1 *}(h) D_{00}^1(h) F(g,h)
          \int_0^{\infty}\! dk k^2
                      \omega_{\pi} \xi(k)^2 \ ,
\nonumber\\
\label{4.5}
\end{eqnarray}
where only $C_{\nu K}$ with $K=0$ appears
because of the axial symmetry of $f_{\mu}(\mathbf{k})$.
The function $F(g,h)$ is defined as
\begin{equation}
  F(g,h) =\exp\left[ s\ D^1_{00}(h) \sum_{\rho'\rho}
               D^1_{\rho'\rho}(g)\ f_{\rho'}f^*_{\rho} \right] \ ,
\label{4.10}
\end{equation}
where the norm integral is given by
\begin{eqnarray}
 s = \int_0^{\infty}\! dk k^2 \xi(k)^2 \ .
\label{4.15}
\end{eqnarray}
We notice that $s$ is not a pripri normalized as $s=1$.
Here we consider the vector $f'_{\mu}$
which is related with $f_{\mu}$ through the Euler angle $g'$ as
$f_{\mu}=\sum_{\nu} D_{\mu\nu}^{1 *}(g') f'_{\nu}$.
With this $f_{\mu}'$, Eq.~(\ref{4.5}) becomes as
\begin{eqnarray}
 (\ref{4.5})&=&
   \sum_{\nu'\nu \lambda'\lambda}
    C'^{*}_{\nu' 0} C'_{\nu 0} f_{\lambda'}^{' *} f'_{\lambda}
\nonumber\\
 &&\times \int\! dg dh
    D_{\nu'\nu}^{1 *}(g) D_{\lambda'\lambda}^{1 *}(g)
    D_{00}^{1 *}(h) D_{00}^1(h) F'(g,h)
    \int_0^{\infty}\! dk k^2 \omega_{\pi} \xi(k)^2 \ ,
\label{4.20}
\end{eqnarray}
where $C'_{\nu 0}= \sum_{\sigma}C_{\sigma 0} D_{\sigma\nu}^1(g')$,
and $F'(g,h)$ is obtained from Eq.~(\ref{4.10})
by exchanging $f_{\mu}$ with $f_{\mu}'$.
Equations (\ref{4.5}) and (\ref{4.20}) show that
the matrix element is invariant with respect to the simultaneous rotation
of $f_{\mu}$ and $C_{\nu 0}$ in the isospace.
Because this is also the case for all other matrix elements,
we take the special value $C_{\nu 0}=(0, 1, 0)$ 
and vary the direction of $f_{\mu}$.

We write the matrix elements of $H$ following Refs.~\cite{Go88,Al99},
\begin{eqnarray}
 \langle N | H | N\rangle = 4\pi\int_0^{\infty}\! dr r^2\left[
      \alpha^2 E_{\alpha\alpha}(r)+\beta^2 E_{\beta\beta}(r)
     +\gamma^2 E_{\gamma\gamma}(r)
   + 2\alpha\beta E_{\alpha\beta}(r)
   + 2\alpha\gamma E_{\alpha\gamma}(r) \right] \ ,
\label{4.25}
\end{eqnarray}
where $E_{\alpha\beta}=E_{\beta\alpha}$, $E_{\alpha\gamma}=E_{\gamma\alpha}$,
and $E_{\beta\gamma}=E_{\gamma\beta}=0$.
The energy densities $E_{ij}$ 
are expressed in terms of the quark fields
and the meson fields $\sigma(r), \phi(r)$ (and $\phi_p(r)$) defined as
\begin{eqnarray}
  \sigma(r) &=& 2 \int_0^{\infty}\!\frac{dk k^2}{\sqrt{\omega_{\sigma}}}
                j_0(kr)\left[\eta(k)+\eta^*(k)\right]\ ,
\label{4.30}\\
  \phi(r) &=& \frac{1}{2\pi}\int_0^{\infty}\!
                   \frac{dk k^2}{\sqrt{\omega_{\pi}}} j_1(kr) \xi(k) \ ,
\label{4.35}\\
  \phi_p(r)&=&\frac{2}{\pi}\int_0^{\infty}\!\! dr' r'^2
                  \int_0^{\infty}\!\! dk k^2
                  \omega_{\pi} j_1(kr) j_1(kr')\ \phi(r') \ .
\label{4.40}
\end{eqnarray}
The diagonal parts are
\begin{equation}
 E_{ii}(r)= E_0(r) n^{ii}_0
          + 2 \phi_p^2\ n^{ii} _1
          + \lambda^2 \left( \sigma^2 - f_{\pi}^2 \right) \phi^2\ n^{ii}_2
          + \frac{\lambda^2}{4} \phi^4\ n^{ii}_3 \ ,
\label{4.45}
\end{equation}
where $i=\alpha, \beta, \gamma$ and
\begin{eqnarray}
 E_0(r) &=& 3\left( 2 u\frac{dv}{dr} + 4\frac{1}{r^2} u v \right)
         + \frac{1}{2}\left( \frac{d\sigma}{dr}\right)^2
         + 3g\sigma \left( u^2 - v^2  \right)
\nonumber\\
        && + \frac{\lambda^2}{4} \left( \sigma^2 - f_{\pi}^2 \right)^2
           + \frac{m_{\pi}^2}{2}\left( \sigma^2 - f_{\pi}^2 \right)
           - m_{\pi}^2 f_{\pi} \left( \sigma - f_{\pi} \right) \ ,
\label{4.50}
\end{eqnarray}
where $n^{ii}_k$ ($k=0,1,2,3$) represent
the integrals with respect to the Euler angles $g$, $h$.
The off-diagonal parts are
\begin{equation}
 E_{\alpha j}(r) = - g\ C_{\alpha j}\ u v\phi\ n^{\alpha j} \ ,
\label{4.55}
\end{equation}
where $j=\beta, \gamma$, and 
$C_{\alpha\beta}=10/ \sqrt{3}$, $C_{\alpha\gamma}= 8\sqrt{2/3}$.
The explicit forms of $n^{ii}_k$ and $n^{\alpha j}$ are summarized in Appendix.

The norm of $|N\rangle$ is expressed in terms of $n^{ii}_0$ as
\begin{equation}
 \langle N | N\rangle =\alpha^2 n^{\alal}_0 +(\beta^2 +\gamma^2) n^{\bebe}_0 \ .
\label{4.65}
\end{equation}
We determine the mixing coefficients $\alpha$, $\beta$, and $\gamma$ 
so that $\langle N|N\rangle=1$.

The differential equations for the quark and meson fields are
\begin{eqnarray}
 \frac{du}{dr} &=&
  -\left( G \sigma + \epsilon \right) v
  -\frac{2}{3} G \alpha \delta_N\ u \phi\ n^{\alpha\beta}
\label{4.70} \ ,\\
 \frac{dv}{dr} &=&
  -\frac{2}{r}v -\left( G \sigma - \epsilon \right) u
  +\frac{2}{3} G \alpha \delta_N\ v \phi\ n^{\alpha\beta}
\label{4.75} \ ,\\
 \frac{d^2\sigma}{dr} &=&
     - \frac{2}{r}\frac{d\sigma}{dr}
     + 3G \left( u^2 - v^2 \right)
     + \lambda^2\left( \sigma^2 - f_{\pi}^2 \right)\sigma
     + m_{\pi}^2\left( \sigma - f_{\pi} \right)
\nonumber\\
   &&+ 2 \lambda^2 \phi^2 \sigma
       \left( \bm{f}\cdot\bm{f} + \frac{1}{9}N_{\pi} \right)
\label{4.80} \ ,\\
 \frac{d^2\phi}{dr^2} &=&
  - \frac{2}{r}\frac{d\phi}{dr} + \frac{2}{r^2}\phi + m_{\pi}^2 \phi
\nonumber\\
 &&+\frac{\lambda^2}{2} \left( \sigma^2 -f_{\pi}^2 \right) \phi
             \left( 1 + \bm{f}\cdot\bm{f} \frac{s}{N_{\pi}} \right)
   +\frac{\lambda^2}{4} \frac{2}{N_{\pi}}
    \left[ \alpha^2 n^{\alal}_3 + ( \beta^2 + \gamma^2 ) n^{\bebe}_3 \right]
    \phi^3
\nonumber\\
 && -G\ \alpha\ \delta_N \ \frac{s}{N_{\pi}}\ n^{\alpha\beta}\ u v
   - E\ \phi_p + \frac{1}{N_{\pi}}\ \phi_p\ 
                 \vartheta(\alpha,\beta,\gamma,u,v,\sigma,\phi,s,f_{\mu}) \ ,
\label{4.85}
\end{eqnarray}
where $\delta_N = (5\beta + 4\sqrt{2}\gamma)/\sqrt{3}$, and
$N_{\pi}=\alpha^2 n^{\alal}_1 + (\beta^2 +\gamma^2) n^{\bebe}_1$
is the expectation value of the pion number operator.
The explicit form of $\vartheta$ is not exhibited here
because it is lengthy but its derivation is straightforward.

The boundary conditions are
\begin{equation}
 \left.\frac{d\sigma}{dr}\right|_{r=0}= 0,\ \ \ v(0)= 0,\ \ \ \phi(0) = 0 \ ,
\label{4.90}
\end{equation}
and for $r \rightarrow \infty$
\begin{eqnarray}
 \left[ r(g^2 f_{\pi}^2 -\epsilon^2)^{1/2}+1 \right]u
                                        -r( g f_{\pi}+\epsilon)v &=&0 \ ,
\nonumber\\
 (2+2m_{\pi}r+m_{\pi}^2 r^2)\phi + (r+m_{\pi}r^2)\frac{d\phi}{dr}&=&0 \ ,
\nonumber\\
 (1+m_{\sigma}r)\left( \sigma - f_{\pi}\right) + r\frac{d\sigma}{dr}&=&0 \ .
\label{4.105}
\end{eqnarray}
We calculate the nucleon mass by solving
the differential equations (\ref{4.70})-(\ref{4.85})
with the boundary conditions (\ref{4.90}), (\ref{4.105})
by the iteration procedure. 

Before finishing this section, 
we comment on the CPA for the pion field \cite{Go88,Al99}. 
We found some errors in Refs.~\cite{Go88,Al99}.
Their treatment of the coherence parameter $x$
as an independent variable is misunderstanding.
Since the dependence on other quantities is dismissed,
the value of $x$ is not correctly determined
in their calculation of the nucleon mass.
In order to compare our result with that of the CPA,
we calculate the nucleon mass in the CPA, too.
\section{\label{sec5} DISCUSSION}
\begin{center}
 a. \ Quasi-coherent state and Coherent pair state
\end{center}

We calculate the nucleon and delta masses.
The pion mass and the decay constant are fixed to the observed values:
$m_{\pi}=140$ MeV, $f_{\pi}=93$ MeV.
The free parameters in our model are
the pion-quark coupling constant $G$ and the sigma mass $m_{\sigma}$.
We choose the typical values for $G$ and $m_{\sigma}$
in order to compare our results directly with those calculated by
using the CPA \cite{Go88,Al99} and the hedgehog ansatz \cite{Fi88}.

Using the parameter set $G=5$, $m_{\sigma}=700$ MeV
 taken from Ref.~\cite{Al99},
we find a self-consistent solution in our model with the QCS
for the pion field.
The quark and meson fields ($u(r)$, $v(r)$, $\phi(r)$, and $\sigma(r)$)
are exhibited in Fig.~\ref{fig1}.
The nucleon mass ($E^{\text{QCS}}_N$) and the delta mass
($E^{\text{CPA}}_{\Delta}$) become 1113 MeV and 1248 MeV, respectively.
\begin{figure}[tbp]
  \includegraphics{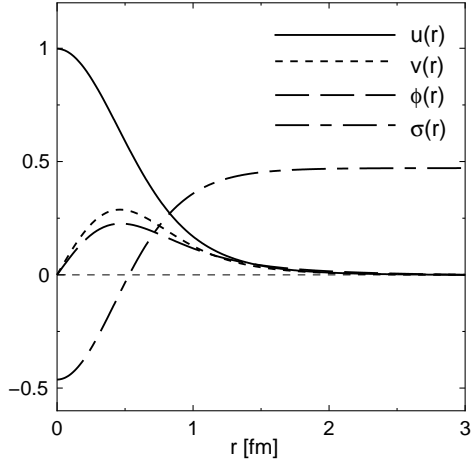}
  \caption{ The quark and meson fields for the nucleon
           using the parameter set $G=5$, $m_{\sigma}=700$ MeV.
           The pion field is described by the QCS.  }
  \label{fig1}
\end{figure}

When we employ the CPA for the pion field instead of the QCS,
we obtain $E^{\text{CPA}}_N = 1093$ MeV and
$E^{\text{CPA}}_{\Delta} = 1233$ MeV.
Because we correct some errors in Refs.~\cite{Go88,Al99},
$E^{\text{CPA}}_N$ is larger than the value found in these references
 by about 20 MeV.

As long as we notice $E_N^{\text{QCS}} >  E_N^{\text{CPA}}$,
the CPA state looks better than the QCS as a trial function
in the variation method.
However, this difference
is only about 2\% of the observed nucleon mass (940 MeV),
and this is also the case for the $\Delta$ mass.
Thus we carefully compare
the QCS with the CPA state.
We can expect similarity in the structure
between these states.

As explained in Sec.~II, the QCS is extracted from the standard coherent state
by using the PY projector both for the isospin and the angular momentum.
We verify a similarity between the CPA state and the QCS
by assuming $f_{\mu}(\mathbf{k}) = f_{\mu}\xi(\mathbf{k})$ as before.
First we consider only the isospin projection
in the definition Eq.~(\ref{2.25}),
\begin{equation}
  |\bm{f};T\mu\nu\rangle
 = P^T_{\mu\nu} \exp \left( \bm{f}\cdot \bm{b}^{\dag} \right) |0\rangle \ ,
\label{5.1}
\end{equation}
where $\bm{b}^{\dag}$ is defined by
\begin{equation}
 b_{\mu}^{\dag}= \int\!d^3k \xi(\mathbf{k})a^{\dag}_{\mu}(\mathbf{k}) \ .
\label{5.5}
\end{equation}
Making use of the Rayleigh expansion, we rewrite Eq.~(\ref{5.1})
further as \cite{Er79}
\begin{equation}
  |\bm{f};T\mu\nu\rangle
  =\sqrt{\frac{4\pi}{2T+1}} 2^T |f|^T Y_{T\nu}^{\ *}(\hat{f})
   \sum_n \frac{ 4\pi 2^T (n+T)! }{ n! (2n+2T+1)! } (\bm{f}\cdot\bm{f})^n
  \left( \bm{b}^{\dag} \cdot \bm{b}^{\dag} \right)^n
   Y_{T\mu}(\bm{b}^{\dag}) |0\rangle \ .
\label{5.10}
\end{equation}
This state satisfies the equation
\begin{equation}
    \bm{b}\cdot \bm{b} |\bm{f};T\mu\nu\rangle
  = \bm{f}\cdot \bm{f} s'^{ 2} |\bm{f};T\mu\nu\rangle \ ,
\label{5.15}
\end{equation}
where $\bm{b}\cdot \bm{b}=\sum_{\mu} (-)^{\mu}b_{\mu}b_{-\mu}$
and $s'=\int\!d^3k \xi(\mathbf{k})^2$.
The QCS is reduced to the CPA state,
if we put the momentum dependence aside from our consideration.

Next the angular momentum is projected,
and the QCS with $T=L=0$ is explicitly written as
\begin{equation}
  |\bm{f};00\rangle \propto \left[
     1 + \frac{1}{18}( \bm{f}\cdot\bm{f} )( \bm{b}^{\dag}\cdot \bm{b}^{\dag} )
       + \cdots \right] |0\rangle \ ,
\label{5.20}
\end{equation}
where 
\begin{equation}
  b^{\dag}_{\mu m}
  = \int_{0}^{\infty}\!{\rm d}k k^2 \xi(k)a^{\dag}_{\mu m}(k) \ ,
\label{5.25}
\end{equation}
and $\bm{b}^{\dag}\cdot\bm{b}^{\dag}
     =\sum(-)^{\mu+m}b^{\dag}_{\mu m}b^{\dag}_{-\mu,-m}$.
The QCS (\ref{5.20}) is no longer the eigenstate of the pion pair
$\bm{b}\cdot\bm{b}$,
and is not equivalent to the CPA state for the pion field in the strict sense.

The first two terms of Eq.~(\ref{5.20}) exactly agree with
those of the expanded form of the CPA pion state \cite{Go88}.
They seem to give the dominant contribution to the nucleon mass
in each model since $\bm{f}\cdot\bm{f}$ and
$\langle \bm{b}^{\dag}\cdot\bm{b}^{\dag}\rangle \sim 1$ in our calculation.
Furthermore the coherence parameter $x$ in Refs.~\cite{Go88,Al99}
corresponds to our $\bm{f}\cdot\bm{f}$.
This situation is also true for the pion state with $T=L=1$.
Thus the structure of the CPA state, which is intuitively defined
in analogy with the standard coherent state, 
can be clearly understood on the basis of the QCS.
Indeed, no essential difference is exposed between the QCS and the CPA state
in our calculation of the nucleon mass.
\begin{center}
 b. \ Quasi-coherent state and Hedgehog ansatz
\end{center}

We take $G=5$ and $m_{\sigma}=1200$ MeV from Ref.~\cite{Fi88},
and obtain $E^{\text{QCS}}_{N}=1215$ MeV.
Employing the hedgehog ansatz for the pion field,
Fiolhais \textit{et al.} obtained $E^{hh}_{N}=938$ MeV \cite{Fi88},
which is lower than our result by about 300 MeV! 
We will show below that the difference between these two models
is in a functional form of the pion distribution. 

It is known that the pion field satisfying the hedgehog ansatz is
closely related to the coherent state \cite{Fi88}.
The expectation value of the field operator
between the standard coherent state $|\Pi\rangle$ is written as
\begin{eqnarray}
 \langle \Pi | \hpi_{\mu}(\mathbf{r}) |\Pi \rangle
 = 2 i \int_0^{\infty}\!\frac{dk\ k^2}{\sqrt{\pi \omega_{\pi}(k)}}
     \sum_{m} j_1(kr)\ {\rm Im} f_{\mu m}(k) Y_{1m}(\hat{r}) \ ,
\label{6.5}
\end{eqnarray}
where only the $p$-wave pion is included as before.
If we assume the correlation between the coordinate and isovector spaces,
 $f_{\mu m}(k) = f(k)\delta_{\mu m}$, Eq.~(\ref{6.5}) becomes
\begin{eqnarray}
 (\ref{6.5})
 = 2 i Y_{1\mu}(\hat{r})
     \int_0^{\infty}\!\frac{dk\ k^2}{\sqrt{\pi \omega_{\pi}(k)}}
     j_1(kr)\ {\rm Im} f(k) \ ,
\label{6.15}
\end{eqnarray}
and we obtain 
$\langle\Pi|\pi_{\mu}(\mathbf{r})|\Pi\rangle=i Y_{1\mu}(\hat{r})\Phi(r)$
 corresponding to the hedgehog ansatz.

A difference is in the topological realization for the pion distribution.
The hedgehog ansatz takes a non-trivial configuration
for the coordinate-isospin mapping with
the winding number 1 in $\pi_3(S_3)=Z$ \cite{Sk61}.
On the other hand, the pion distribution in our QCS is topologically trivial,
i.e. the winding number 0.
This mathematical difference in geometrical character 
has significant effect on our physical interpretation 
of the nucleon structure obtained in the models.
The mass difference should not be taken
merely as a matter of choice of a trial function
in the variational method.

The above discussion shows that a pure imaginary function
 may be chosen to the pion distribution if the hedgehog ansatz is considered.
In our model, however, the pion distribution takes a simple real form.
We can show that our equations are independent of
the phase of a complex function $f_{\mu}(\bm{k})$
as far as the baryon masses are considered.
Note that
the expectation value of $\hpi_{\mu}(\mathbf{r})$ for the QCS 
is not proportional to Im$f_{\mu m}(k)$
because of the PY projection on the pion coherent state.

We consider that the 300 MeV-difference in the nucleon mass
is related with the topological problem.
This observation leads us to the following interpretation on our solution.
Insofar as the ground state properties are concerned,
we can accept that the hedgehog pion may be suitable for the nucleon.
The pion distribution in the QCS is
topologically distinguished from that in the hedgehog ansatz,
and our solution may correspond to the excited state of the nucleon.
This excitation is caused by the change in the pion configuration,
which is completely different from the usual mechanism of baryon excitation
in a constituent quark model.
We know that some kinds of the nucleon resonances, such as the Roper resonance,
are not fully explained by the quark excitation.
We need to solve this problem, for example, by introducing new 
degrees of freedom other than the constituent quarks.
The novel structure of our solution may be taken as one of the possible
mechanisms for the baryon excitation.
\section{\label{sec6}SUMMARY AND PERSPECTIVES}

We have calculated the nucleon mass in the linear sigma model
describing the pion field by using the standard coherent state.
The trivial topology is chosen for the pion field,
and the QCS is constructed by using the PY projection.
We can understand the CPA on the basis of the QCS.
We have also shown that the topological difference between the QCS
and the hedgehog state
is important in the nucleon mass.

Now we comment on a critical issue in the application
of the standard coherent state to the baryon physics.
As we discussed in this work, the standard coherent state is suitable
for the comprehensive studies of the static pion field in a baryon.
As for the excited baryons, the pion spatial excitation must be
taken into account in addition to the quark excitation,
which is not a serious problem when the ground state properties are considered.
In the nonlinear theory of scalar particles, we usually quantize
the fluctuation around the static field with minimum energy.
However, the isospin symmetry makes this quantization procedure
significantly difficult for the standard coherent state.
Although the projection method is often employed before proceeding to
the spatial quantization, this approach does not actually work 
because the excitation energy found in this projection
is on the same order of the spatial excitation energy \cite{Br85,Bl88}.

We are now seeking for the general method of constructing a pion state
without relying on the standard coherent state in the nonlinear problem.
Generalization of the coherent state based on the group theory
may be a possible clue to tackle this problem \cite{Pe86}.
\appendix*
\section{}

The explicit forms of $n^{ij}_k$ ($i,j=\alpha,\beta,\gamma$ and $k=0,1,2,3$)
in the energy densities are exhibited.
For $n^{\alal}_k$,
\begin{eqnarray}
 n^{\alal}_0&=& \int\!{\rm d}g {\rm d}h F(g,h) \ ,
\label{app.1}\\
 n^{\alal}_1 &=&
  \sum_{\lambda'\lambda} f_{\lambda'}f_{\lambda}^*
             \int\! dg dh D^1_{00}(h) D^1_{\lambda'\lambda}(g) F(g,h)
\label{app.5}\\
 n^{\alal}_2 &=& \bm{f}\cdot\bm{f}\ n^{\alal}_0 + n^{\alal}_1 \ ,
\label{app.10}\\
 n^{\alal}_3 &=& \frac{4}{5} (\bm{f}\cdot\bm{f})^2
        \left[ 7 n^{\alal}_0 + 2 \int\! dg dh D^2_{00}(h) F(g,h) \right]
      + \frac{72}{5} \bm{f}\cdot\bm{f} n^{\alal}_1
\nonumber\\
    &+& 4 \sum_{\lambda'\lambda\tau'\tau}
        f_{\lambda'} f_{\lambda}^* f_{\tau'} f_{\tau}^*
         \int\!{\rm d}g {\rm d}h \left[ 1 + \frac{4}{5}D^2_{00}(h) \right]
                  D^1_{\lambda'\lambda}(g) D^1_{\tau'\tau}(g) F(g,h) \ ,
\nonumber\\
\label{app.15}
\end{eqnarray}
where $s$ and $F(g,h)$ are defined in the text.

We can obtain $n^{\bebe}_k=n^{\gaga}_k$ by 
inserting $1/9\times D^1_{00}(h)D^1_{00}(g)$ in $n^{\alal}_k$.
For example,
\begin{equation}
  n^{\bebe}_0 
  = \frac{1}{9}\int\! dg dh D^1_{00}(h) D^1_{00}(g) F(g,h) \ .
\label{app.25}
\end{equation}
Equations (\ref{app.1}) and (\ref{app.25}) are the norms of
the QCS $|\bm{f};00\rangle$ and $|\bm{f};11\rangle$, respectively.

In the off-diagonal densities, $n^{\alpha\beta}=n^{\alpha\gamma}$, and
\begin{equation}
  n^{\alpha\beta} = \frac{1}{9} \left[ \bm{f}\cdot\bm{f}
             + \sum_{\sigma} f_{\sigma}
  \int\! dg dh D^1_{00}(h) D^1_{\sigma 0}(g) F(g,h) \right] \ .
\label{app.30}
\end{equation}
\begin{acknowledgments}
The authors thank
Prof.~T.~Sato and Prof.~M.~Wakamatsu and Mr.~H.~Kamano for useful discussions.
\end{acknowledgments}
%

\begin{thebibliography}{18}
\expandafter\ifx\csname natexlab\endcsname\relax\def\natexlab#1{#1}\fi
\expandafter\ifx\csname bibnamefont\endcsname\relax
  \def\bibnamefont#1{#1}\fi
\expandafter\ifx\csname bibfnamefont\endcsname\relax
  \def\bibfnamefont#1{#1}\fi
\expandafter\ifx\csname citenamefont\endcsname\relax
  \def\citenamefont#1{#1}\fi
\expandafter\ifx\csname url\endcsname\relax
  \def\url#1{\texttt{#1}}\fi
\expandafter\ifx\csname urlprefix\endcsname\relax\def\urlprefix{URL }\fi
\providecommand{\bibinfo}[2]{#2}
\providecommand{\eprint}[2][]{\url{#2}}

\bibitem[{\citenamefont{Tomonaga}(1947)}]{To47}
\bibinfo{author}{\bibfnamefont{S.}~\bibnamefont{Tomonaga}},
  \bibinfo{journal}{Prog. Theor. Phys.} \textbf{\bibinfo{volume}{2}},
  \bibinfo{pages}{6} (\bibinfo{year}{1947}).

\bibitem[{\citenamefont{Lee and Pines}(1953)}]{Le53}
\bibinfo{author}{\bibfnamefont{T.~D.} \bibnamefont{Lee}} \bibnamefont{and}
  \bibinfo{author}{\bibfnamefont{D.}~\bibnamefont{Pines}},
  \bibinfo{journal}{Phys.\ Rev.} \textbf{\bibinfo{volume}{92}},
  \bibinfo{pages}{883} (\bibinfo{year}{1953}).

\bibitem[{\citenamefont{Skyrme}(1961)}]{Sk61}
\bibinfo{author}{\bibfnamefont{T.~H.~R.} \bibnamefont{Skyrme}},
  \bibinfo{journal}{Proc.\ R.\ Soc.\ London} \textbf{\bibinfo{volume}{A260}},
  \bibinfo{pages}{127} (\bibinfo{year}{1961}).

\bibitem[{\citenamefont{Fiolhais \textit{et~al.}}(1988)\citenamefont{Fiolhais,
  Goeke, Gr{\"u}mmer, and Urbano}}]{Fi88}
\bibinfo{author}{\bibfnamefont{M.}~\bibnamefont{Fiolhais}},
  \bibinfo{author}{\bibfnamefont{K.}~\bibnamefont{Goeke}},
  \bibinfo{author}{\bibfnamefont{F.}~\bibnamefont{Gr{\"u}mmer}},
  \bibnamefont{and} \bibinfo{author}{\bibfnamefont{J.~N.}
  \bibnamefont{Urbano}}, \bibinfo{journal}{Nucl. Phys.}
  \textbf{\bibinfo{volume}{A481}}, \bibinfo{pages}{727} (\bibinfo{year}{1988}).

\bibitem[{\citenamefont{Bolsterli}(1981)}]{Bo81}
\bibinfo{author}{\bibfnamefont{M.}~\bibnamefont{Bolsterli}},
  \bibinfo{journal}{Phys.\ Rev.\ D} \textbf{\bibinfo{volume}{24}},
  \bibinfo{pages}{400} (\bibinfo{year}{1981}).

\bibitem[{\citenamefont{Goeke \textit{et~al.}}(1988)\citenamefont{Goeke,
  Harvey, Gr{\"u}mmer, and Urbano}}]{Go88}
\bibinfo{author}{\bibfnamefont{K.}~\bibnamefont{Goeke}},
  \bibinfo{author}{\bibfnamefont{M.}~\bibnamefont{Harvey}},
  \bibinfo{author}{\bibfnamefont{F.}~\bibnamefont{Gr{\"u}mmer}},
  \bibnamefont{and} \bibinfo{author}{\bibfnamefont{J.~N.}
  \bibnamefont{Urbano}}, \bibinfo{journal}{Phys.\ Rev.\ D}
  \textbf{\bibinfo{volume}{37}}, \bibinfo{pages}{754} (\bibinfo{year}{1988}).

\bibitem[{\citenamefont{Adkins \textit{et~al.}}(1983)\citenamefont{Adkins,
  Nappi, and Witten}}]{Ad83}
\bibinfo{author}{\bibfnamefont{G.~S.} \bibnamefont{Adkins}},
  \bibinfo{author}{\bibfnamefont{C.~R.} \bibnamefont{Nappi}}, \bibnamefont{and}
  \bibinfo{author}{\bibfnamefont{E.}~\bibnamefont{Witten}},
  \bibinfo{journal}{Nucl. Phys.} \textbf{\bibinfo{volume}{B228}},
  \bibinfo{pages}{552} (\bibinfo{year}{1983}).

\bibitem[{\citenamefont{Wakamatsu}(1992)}]{Wa92}
\bibinfo{author}{\bibfnamefont{M.}~\bibnamefont{Wakamatsu}},
  \bibinfo{journal}{Prog.\ Theor.\ Phys. Supp} \textbf{\bibinfo{volume}{109}},
  \bibinfo{pages}{115} (\bibinfo{year}{1992}).

\bibitem[{\citenamefont{Eriksson \textit{et~al.}}(1981)\citenamefont{Eriksson,
  Mukunda, and Skagerstam}}]{Er81}
\bibinfo{author}{\bibfnamefont{K.-E.} \bibnamefont{Eriksson}},
  \bibinfo{author}{\bibfnamefont{N.}~\bibnamefont{Mukunda}}, \bibnamefont{and}
  \bibinfo{author}{\bibfnamefont{B.-S.} \bibnamefont{Skagerstam}},
  \bibinfo{journal}{Phys.\ Rev.\ D} \textbf{\bibinfo{volume}{24}},
  \bibinfo{pages}{2615} (\bibinfo{year}{1981}).

\bibitem[{\citenamefont{Peierls and Yoccoz}(1957)}]{Pe57}
\bibinfo{author}{\bibfnamefont{R.~E.} \bibnamefont{Peierls}} \bibnamefont{and}
  \bibinfo{author}{\bibfnamefont{J.}~\bibnamefont{Yoccoz}},
  \bibinfo{journal}{Roy.\ Soc.\ London A} \textbf{\bibinfo{volume}{70}},
  \bibinfo{pages}{381} (\bibinfo{year}{1957}).

\bibitem[{\citenamefont{Aly \textit{et~al.}}(1999)\citenamefont{Aly, McNeil,
  and Pruess}}]{Al99}
\bibinfo{author}{\bibfnamefont{T.~S.~T.} \bibnamefont{Aly}},
  \bibinfo{author}{\bibfnamefont{J.~A.} \bibnamefont{McNeil}},
  \bibnamefont{and} \bibinfo{author}{\bibfnamefont{S.}~\bibnamefont{Pruess}},
  \bibinfo{journal}{Phys.\ Rev.\ D} \textbf{\bibinfo{volume}{60}},
  \bibinfo{pages}{114022} (\bibinfo{year}{1999}).

\bibitem[{\citenamefont{Rose}(1957)}]{Ro57}
\bibinfo{author}{\bibfnamefont{M.~E.} \bibnamefont{Rose}},
  \emph{\bibinfo{title}{Elementary Theory of Angular Momentum}}
  (\bibinfo{publisher}{John Wiley \& Sons}, \bibinfo{address}{New York},
  \bibinfo{year}{1957}).

\bibitem[{\citenamefont{Ring and Schuck}(1980)}]{Ri80}
\bibinfo{author}{\bibfnamefont{P.}~\bibnamefont{Ring}} \bibnamefont{and}
  \bibinfo{author}{\bibfnamefont{P.}~\bibnamefont{Schuck}},
  \emph{\bibinfo{title}{The Nuclear Many-Body Problems}}
  (\bibinfo{publisher}{Springer-Verlag}, \bibinfo{address}{New York},
  \bibinfo{year}{1980}).

\bibitem[{\citenamefont{Isgur and Karl}(1979)}]{Is79}
\bibinfo{author}{\bibfnamefont{N.}~\bibnamefont{Isgur}} \bibnamefont{and}
  \bibinfo{author}{\bibfnamefont{G.}~\bibnamefont{Karl}},
  \bibinfo{journal}{Phys.\ Rev.\ D} \textbf{\bibinfo{volume}{19}},
  \bibinfo{pages}{2653} (\bibinfo{year}{1979}).

\bibitem[{\citenamefont{Eriksson and Skagerstam}(1979)}]{Er79}
\bibinfo{author}{\bibfnamefont{K.-E.} \bibnamefont{Eriksson}} \bibnamefont{and}
  \bibinfo{author}{\bibfnamefont{B.-S.} \bibnamefont{Skagerstam}},
  \bibinfo{journal}{J.\ Phys.\ A} \textbf{\bibinfo{volume}{12}},
  \bibinfo{pages}{2175} (\bibinfo{year}{1979}).

\bibitem[{\citenamefont{Braaten and Ralston}(1985)}]{Br85}
\bibinfo{author}{\bibfnamefont{E.}~\bibnamefont{Braaten}} \bibnamefont{and}
  \bibinfo{author}{\bibfnamefont{J.~P.} \bibnamefont{Ralston}},
  \bibinfo{journal}{Phys.\ Rev.\ D} \textbf{\bibinfo{volume}{31}},
  \bibinfo{pages}{598} (\bibinfo{year}{1985}).

\bibitem[{\citenamefont{Blaizot and Ripka}(1988)}]{Bl88}
\bibinfo{author}{\bibfnamefont{J.~P.} \bibnamefont{Blaizot}} \bibnamefont{and}
  \bibinfo{author}{\bibfnamefont{G.}~\bibnamefont{Ripka}},
  \bibinfo{journal}{Phys.\ Rev.\ D} \textbf{\bibinfo{volume}{38}},
  \bibinfo{pages}{1556} (\bibinfo{year}{1988}).

\bibitem[{\citenamefont{Perelomov}(1986)}]{Pe86}
\bibinfo{author}{\bibfnamefont{A.}~\bibnamefont{Perelomov}},
  \emph{\bibinfo{title}{Generalized Coherent States and Their Applications}}
  (\bibinfo{publisher}{Springer-Verlag}, \bibinfo{address}{Berlin},
  \bibinfo{year}{1986}).

\end{thebibliography}
%

%
\end{document}